OPEN

# CHD Risk Minimization through Lifestyle Control: Machine Learning Gateway

Xi He[1], B. Rajeswari Matam[1], Srikanth Bellary[2,3], Goutam Ghosh[4] & Amit K. Chattopadhyay[1*]

Studies on the influence of a modern lifestyle in abetting Coronary Heart Diseases (CHD) have mostly focused on deterrent health factors, like smoking, alcohol intake, cheese consumption and average systolic blood pressure, largely disregarding the impact of a healthy lifestyle in mitigating CHD risk. In this study, 30+ years' World Health Organization (WHO) data have been analyzed, using a wide array of advanced Machine Learning techniques, to quantify how regulated reliance on positive health indicators, e.g. fruits/vegetables, cereals can offset CHD risk factors over a period of time. Our research ranks the impact of the negative outliers on CHD and then quantifies the impact of the positive health factors in mitigating the negative risk-factors. Our research outcomes, presented through simple mathematical equations, outline the best CHD prevention strategy using lifestyle control only. We show that a 20% increase in the intake of fruit/vegetable leads to 3–6% decrease in SBP; or, a 10% increase in cereal intake lowers SBP by 3%; a simultaneous increase of 10% in fruit-vegetable can further offset the effects of SBP by 6%. Our analysis establishes gender independence of lifestyle on CHD, refuting long held assumptions and unqualified beliefs. We show that CHD risk can be lowered with incremental changes in lifestyle and diet, e.g. fruit-vegetable intake ameliorating effects of alcohol-smoking-fatty food. Our multivariate data model also estimates functional relationships amongst lifestyle factors that can potentially redefine the diagnostics of Framingham score-based CHD-prediction.

Globally, cardiovascular diseases account for nearly 17.9 million deaths with Coronary Heart Disease (CHD) accounting for 80% of these[1]. A myriad of factors have been identified as risk generators, including ethnicity, sex, total cholesterol level, triglycerides, blood pressure, that in turn are affected by life style denominators[2,3]. Together, they determine the risk appraisal function that have been assessed using conventional predictive scoring like body mass index (BMI) and Framingham scores[4,5] together with more advanced population biology or epidemiological estimators. Although there have been numerous advances in the treatment of established CHD, at a population level, assessed through Artificial Intelligence (Machine Learning) based adaptation of established statistical wisdom, remains a major knowledge gap[5].

Ground breaking epidemiological studies have identified key lifestyle and health indicators as risk factors for Coronary Heart Disease (CHD)[6–9]. Lifestyle factors include smoking[10,11], alcohol consumption[12], lack of physical activity while key health indicators include obesity, high blood pressure[5] and diabetes[13]. Evidence suggests a diet rich in fruits, vegetables and whole grains can mitigate the onset of CHD[8]. Some of these risk factors are not individually causative but when combined with other risk factors, increase the risk of CHD[10–14]. Further, lifestyle factors can be suitably modified to ameliorate the severity of CHD but the quantitative relationship between various life style factors and how they impact precise predictions on CHD risk or life expectancy have not been explored properly.

Risk prediction models, such as the Framingham Risk Score (FRS)[15–17]. Systematic Coronary Risk Evaluation (SCORE)[18] and World Health Organization/International Society of Hypertension (WHO/ISH)[17,18] are routinely used in clinical practice to ascertain CHD risk. Many of these models have been validated in specific populations and in smaller cohorts raising concerns that they may overestimate the risk when applied universally. Moreover, some of these models[16] use the individual's lifestyle and general health, including parameters such as age, gender, total (also LDL/HDL) cholesterol, smoking habit and systolic blood pressure (SBP). Each parameter is scored and the sum of all scores, representing the combined impact of all the parameters, identifies the risk. These scores are not necessarily weighted, nor is any mutual correlation necessarily calibrated. For example, smoking and alcohol consumption are known to affect SBP[11,18] but by how much remains uncertain. Similarly, age is a major risk factor for CHD, affecting blood pressure and cholesterol levels, yet the extent of its influence has not been analyzed clearly. This points to a knowledge-gap in identifying how lifestyle factors affect each other, and are further affected by other factors like age, ethnicity and gender, in deciding CHD risk.

[1]Aston University, Systems Analytics Research Institute, Mathematics, Birmingham, B4 7ET, UK. [2]University Hospitals Birmingham NHS Foundation Trust, Birmingham, UK. [3]Aston University, School of Life and Health Sciences, Aston Triangle, Birmingham, B4 7ET, UK. [4]Vice Chancellor, GIET University, Gunupur, 765022, Dt. Rayagada, Odisha, India. *email: a.k.chattopadhyay@aston.ac.uk

| | Specification | Source of data | Collected years |
|---|---|---|---|
| CHD death rate | Calculated by: SDR7 of ischemic heart disease/SDR of all causes, including all ages, per 100 000, males and females separately. | WHO Global Health Observatory Data Repository, available from http://apps.who.int/ghodata/ | 1970–2014 |
| Alcohol Consumption | Recorded adult (15+ years) per capita (APC) consumption of pure alcohol. In order to make the conversion into litres of pure alcohol, the alcohol content of beer, wine and spirits is considered to be 5%, 12% and 40% respectively. Alcohol consumption here includes all alcoholic drinks. | WHO Global Health Observatory Data Repository, available from http://www.who.int/substance_abuse/publications/global_alcohol_report/en/index.html | 1970–2013 |
| Cheese Consumption | Total supply amount of cheese per capita per year in kg. | FAO Statistics Division, Food Supply Sheets, available from http://faostat3.fao.org/home/index.html | 1970–2013 |
| Smoking | Measured using the standard questionnaire during a health interview of a representative sample of the population aged 15 years and above. Data are largely collected from multiple sources by the Tobacco or Health unit at WHO/EURO, males and females separately. | WHO Global Health Observatory Data Repository, available from http://apps.who.int/ghodata/ | 1980–2013 |
| Systolic Blood Pressure | Mean systolic blood pressure trends, age-standardized (mmHg), males and females separately. | WHO Global Health Observatory Data Repository, available from http://apps.who.int/gho/data/view.main.12467EST?lang=en | 1980–2013 |
| Cereal Consumption | Total average amount of cereal available per person per year in kg, excluding beer. | FAO Statistics Division, Food Balance Sheets, available from http://faostat3.fao.org/home/index.html | 1970–2013 |
| Fruits and Vegetables Consumption | Total average amount of fruits and vegetables available per person per year in kg, excluding wine. | FAO Statistics Division, Food Balance Sheets, available from http://faostat3.fao.org/home/index.html | 1970–2013 |

**Table 1.** Detailed data specifications.

Between 2007 and 2017, the impact of life-style (gender difference) risk factors on cardiovascular diseases for 13 EU countries and the USA have been published by WHO[19] together with 12 peer-reviewed research articles. Using population-based survey of people aged 40–84, 5 of these reported inconclusive relationship between lifestyle and CHD mortality; 3 reviewed CHD prevention in men/women but did not clarify life-style impact; only 1 article studied the effect of dairy-fat intake on CHD while 2 others studied the impact of gender difference on CHD; overall these publications essayed largely non-quantified observations, focusing on qualitative impact only. Only one study used mathematical modelling[20] but relied on numerous non-specified assumptions. In general, the conclusion that we may draw is that the findings of these studies do not fully clarify the role of positive life style factors in offsetting the excess CHD risk from adverse risk factors.

The aim of this study is to systematically analyze the causal relationship between lifestyle factors, health indicators and diet. The results obtained are utilized to provide an easy to use advisory tool that can be used by clinicians to evaluate an individual's perceptible risk of CHD and inform the individual as to the necessary changes required to neutralize, or at least, contain adverse health risks.

## Subjects and Methods

**Data collection and modeling dynamics of CHD.** In our project, we used UK data showing standardized prevalence rates for those aged 18 and above per 100,000 persons for the variables smoking, SBP, consumption of alcohol, cheese, fruit and vegetables, cereals or whole grains and deaths attributable to CHD (Tables 1 and 2). Detailed data specification and sources are shown in Table 1 and their relative order of importance, based on Principal Component Analysis, in Table 2.

This is a data modeling study using advanced machine learning tools to quantify the inter-dependence of key life style factors, both positive and negative, in estimating CHD risk. This study provides precise numbers as to how many portions a day could reduce or eliminate CHD risk. This is the first multivariate study that analyzes the impact of multiple life style effects in a real situation when all may vary simultaneously.

Data used in this work were all anonymously collected for the UK male and female populations, between 1990 and 2013, as detailed in Table. Data used were all collected by others from different studies independent of the present authors. The focus on the post-1990 period is indicative of a number of major health initiatives for treatment of CHD globally including control of blood pressure and introduction of statins (cholesterol reducing drug) which were accessible to the UK population through the National Health Service (NHS)[21]. Health statistics show a marked change in the average SBP and number of deaths due to CHD in most of the developed countries from 1990 onwards. In the pre-1990 period, whereas CHD statistics showed a positive gradient of increase with time, the post-1990 era shows a negative gradient. Interestingly, both the pre- and post-1990 statistics abide linear regression profiles, the cause of which is unclear. As such, we used 1990 as the normalization year beyond which significant health risks need to be estimated.

| PCA Component | Eigenvalues - Male | Eigenvalues - Female |
|---|---|---|
| 1 | 5·9512 | 5·9517 |
| 2 | 0·0486 | 0·0483 |
| 3 | 0·0010 | 0·0009 |
| 4 | 0·0006 | 0·0006 |
| 5 | 0·0003 | 0·0003 |
| 6 | 0·0002 | 0·0001 |
| Total | 6·0020 | 6·0020 |

**Table 2.** Grading all CHD variables using Principal Component Analysis (PCA).

Based on large scale population studies[6–9], we divided the data into two causative sets connected by a singular effect, the number of deaths due to CHD. Smoking, SBP, alcohol and cheese consumption were classed as negative indicators, associated with increased risk of CHD. Consumption of fruit and vegetables, and cereals or whole grains were classed as positive indicators, potentially mitigating CHD risk.

Data analysis was conducted in two steps: 1) multivariate analysis of all negative indicators (alcohol, cheese, smoking, SBP) were conducted, keeping the positive indicators ("good" factors, cereal and fruit-vegetable) at their mean (time averaged) values; 2) data variation for all 6 variables were jointly considered. The first step ensured that fluctuations in positive indicators did not obscure or skew the risk factors on CHD while the second step was needed to predict how positive indicators mitigate negative ones.

The statistics and machine learning toolbox of Matlab 2016a 9.0 (verified also with Matlab 2018b 9.5) was used for the data analysis. The available data for each variable was initially analyzed by applying linear regression. This surprisingly showed a linearly decreasing trend for CHD death rate, smoking, Systolic Blood Pressure (SBP), alcohol consumption through the years and a linearly increasing trend for cereal, fruit and vegetable consumption. This enabled the use of a simple mathematical equation $y = a + bx$ where 'x' represents the year and 'y' the value of the missing data for the given variable e.g. smoking, SBP, etc., to Supplement missing data. Regression-based extrapolation over a 25-year range (1990–2013) was then used to Supplement data for specific years synthetically, resulting in a statistically large dataset confirming linear evolution.

Principal Component Analysis (PCA)[22] is the most commonly used statistical method for multivariate data analysis. PCA transforms a given set of possibly correlated variables into a set of uncorrelated variables known as principal components (PCs). The PCs are ordered and represent the variables in order of their dominance in the dataset. PCA of 4 (risk factors only) and 6 (risk and positive factors) dimensional datasets were both mapped on to weighted 2-dimensional hyperspaces, respectively for males and females. All data availed are believed to conform to the relevant guidelines and regulations strictly adhered under the WHO guidelines[1].

Data used in this study were collected from WHO and FAO repositories, as detailed in Table 1, abiding relevant guidelines pertaining to human data collection ethics and regulations. Further details are available from the World Health Organization websites (Table 1).

PCA was then used to rank all affecting life style factors in order of importance (largest eigenvalue implying most important). Results were independently ratified using nonlinear Neuroscale (NSC), Generative Topological Mapping (GTM) and Gaussian Process Latent Variable Model (GPLVM)[22–24] techniques, all essentially converging to the same linear regression formulae.

**Multivariate correlation analysis of all 6 "risk" factors.** For the multivariate risk prediction model, we combined all 6 "risk" factors, negative - SBP, Smoking, Cheese, Alcohol, and positive - cereals and fruit-vegetables, separately for males and females. This was followed by an extended PCA, results verified by NSC, GTM and GPLVM, to analyze the extent of influence of the positive indicators upon the risk factors.

$$SBP_{male} = -0 \cdot 459 \times Alcohol - 0 \cdot 441 \times Smoking - 0 \cdot 442 \\ \times Cheese - 0 \cdot 353 \times Cereals - 0 \cdot 352 \times Fruit-Veg + 0 \cdot 874 \quad (1)$$

$$SBP_{female} = -0 \cdot 454 \times Alcohol - 0 \cdot 434 \times Smoking - 0 \cdot 448 \\ \times Cheese - 0 \cdot 36 \times Cereals - 0 \cdot 359 \times Fruit-Veg + 1 \cdot 273 \quad (2)$$

**Model validation.** In order to validate these results, the entire UK dataset was visualised using all four visualization methods (PCA, NSC, GTM and GPLVM)[23], by applying the following visualization quality evaluation measures:

  a. Trustworthiness
     In order to analyze the model in the reduced dimensionality space, the relatively distant data points were projected in a predefined neighborhood. Our results were then verified against *Trustworthiness* that is estimated as a fraction of data points distant in the original data space that eventually lie within the defined mapped neighborhood.

For a set with cardinality n, if R(i, j) is the rank of the data points j measured against the corresponding data points i with respect to the distance measure in the original data space, and Uk(i) denote the data points in the k-nearest neighborhood of the i data points in the latent visualization space but not in the original data space, Trustworthiness over a subset of k-neighbors can be measured as

$$T(k) = 1 - \frac{2}{nk(2n-3k-1)}\sum_{i=1}^{n}\sum_{j\in U_k(i)}(R(i,j) - k)$$

where nk(2n-3k-1) is the normalizing factor, ensuring the value of trustworthiness stays between 0 and 1; the higher the value, better is the visualization result.

b. Continuity

Continuity is measured as the fraction of neighbouring data points in the original data space that becomes distant in the mapping space. If n represents the data size, R(i,j) the rank of the data points, j a running index scanning the corresponding data points, i a dummy index representing the distance measured in the latent visualization space, and Vk(i) denote the data points in the k-nearest neighbourhood of the i data points in the original data space but not in the latent visualization space, Trustworthiness with k-neighbors can be measured as

$$C(k) = 1 - \frac{2}{nk(2n-3k-1)}\sum_{i=1}^{n}\sum_{j\in V_k(i)}(R^*(i,j) - k)$$

Again, nk(2n-3k-1) is the normalizing factor, ensuring the value of continuity is constrained between 0 and 1; the higher the value better is the visualization result.

c. Mean Relative Rank Errors (MRRE)

Mean relative rank errors with respect to data space (MRREd) and latent visualization space (MRREl) is another well-known quality measure that was used in this study. The mean relative rank errors with respect to data space (MRREd) can be calculated from the formula

$$MRREd(k) = \frac{1}{n\sum_{k'=1}^{k}\frac{|n-2k'|}{k'}}\sum_{i=1}^{n}\sum_{j\in U_k(i)}\frac{|(R^*(i,j) - R(i,j)|}{R(i,j)}$$

where the mean relative rank errors with respect to latent visualization space (MRREl) can be calculated by

$$MRREl(k) = \frac{1}{n\sum_{k'=1}^{k}\frac{|n-2k'|}{k'}}\sum_{i=1}^{n}\sum_{j\in n_{k*}(i)}\frac{|(R(i,j) - R_*(i,j)|}{R_*(i,j)}$$

where $n\sum_{k'=1}^{k}\frac{|n-2k'|}{k'}$ is the normalising factor, 0 < MRREl(k) < 1; the lower the value of MRREl, better is the visualization result.

All 3 measures combined lead to the multivariate *quality matrix* shown in Table 3 below.

The results show that for the trained PCA visualization model based on UK datasets, the trustworthiness value is 0.9974 for males and 0.9911 for females; NSC visualization model shows a similar trustworthiness value as PCA, 0.9969 for males and 0.9922 for females; GTM and GPLVM trained with comparative lower trustworthiness value, which is all around 0.9 only. Continuity based validation is equally reciprocal: PCA's continuity values are 0.9969 for males and 0.9953 for females, NSC shows identical values as PCA again. GTM shows lowest continuity value, both for males and females, and are smaller than 0.9 compared to <0.95 after GPLVM training; MRREd and MRREl values are found to be four times higher on GTM and GPLVM training than PCA and NSC, for both males and females.

As trustworthiness and based validation imply higher the value, better the visualization is, while the trends are reversed for MRREd and MRREl, we conclude that PCA and NSC provide better visualization structures for our study. What is significant is that the validation results are not much different between PCA (linear) and NSC (nonlinear). We therefore resorted to primary PCA based evaluation, followed by NSC ratification, at all steps of this study.

## Results

Figure 1a shows the CHD death rate in the UK between 1970–2013, in the pre and post statin era; Fig. 1b on the other hand shows the timeline trend of CHD between 1990–2013 in the UK, separately for men and women. Over a 23-year period, the CHD rate decreased by ca 48% (0·2927 to 0·1529) in men and ca 58% (0·2204 to 0·0930) in women; in the same period, smoking habits reduced by 29% (0·31 to 0·22) in men. Notably, CHD fraction in men is numerically higher than in women, data conforming closely to linear regression fits for both genders.

During the same period, varying trends were observed in the different positive and negative risk factors. Alcohol consumption increased per capita by 22.7% (9·96 Litrers in 1990 to a peak of 12·22 Litres in 2004), followed by a small decrease in 2013. Cheese consumption on the other hand showed a sharp per capita increase by 46.4% (7·74 kg in 1990 to 11·33 kg in 2013). In contrast, both smoking and SBP showed steady decline; male smoking dropped from 29·03% to 22% compared to 41·38% to 31% for female smokers, both males and females have nearly perfect correlations with degrees of freedom and P-values given respectively as follows: $F_{male}$ (1,22) = 166·216, P-value$_{male}$ < 0·0001; $F_{female}$ (1,22) = 351·704, P-value$_{female}$ < 0·0001; comparative SBP levels showed a reduction of 6·3 mm Hg in men and 8·9 mmHg in women within the same period (Fig. 2).

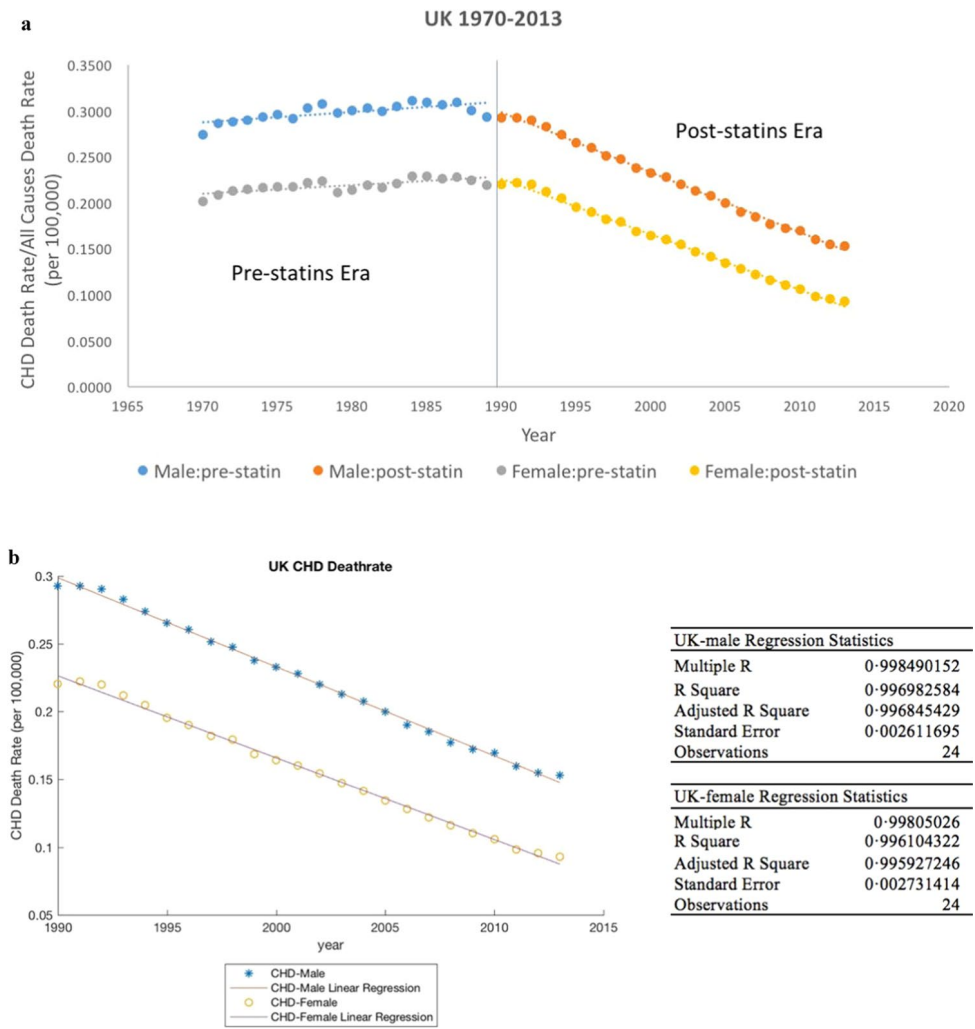

**Figure 1.** (**a**) UK CHD death rate statistics in the pre and post statin era. (**b**) Linear plot and regression fit of CHD death rate.

Tables 3 and 4 compare visualization results between the negative risk factors only (Table 3) against the palliative effects of positive risk factors together with negative factors (Table 4). PC1 and PC2 rank the descriptors. A larger PC1-value indicates greater significance. For PC1-values identical up to the third decimal point, PC2 ranks the descriptors.

Table 4 summarizes the following ranking of the 4-risk factor set, in order of their importance:

Males: SBP > Smoking > Cheese > Alcohol; Females: Smoking > SBP > Cheese > Alcohol.

Table 5 summarizes the following ranking of the 6-risk factor set, as follows:

Males: SBP > Smoking > Cheese > Fruit-Veg > Cereal > Alcohol;
Females: Smoking > SBP > Cheese > Fruit-Veg > Cereal > Alcohol.

The linear relationship obtained between an increase in the risk factors and CHD is consistent with other published studies[20,21]. Table 3 confirms the positive impact of fruit-vegetables on CHD risk. Tables 6 and 7 below (numbers resulting from Eq. (1)) estimates exactly by how much:

**Key outcomes.** *Improvement of framingham description.* Our multivariate data model estimates the nature of functional relationship between the risk and positive lifestyle factors instead of assuming them to be linear.

*Gender inequality estimated.* Amongst men, SBP had the greatest effect followed by smoking, cheese consumption and alcohol respectively, while in women it was smoking that had the biggest impact followed by SBP, cheese consumption and alcohol intake. When all 6 factors were considered together (Table 5), a clear mitigating effect was seen with cereal and fruit-vegetable intake.

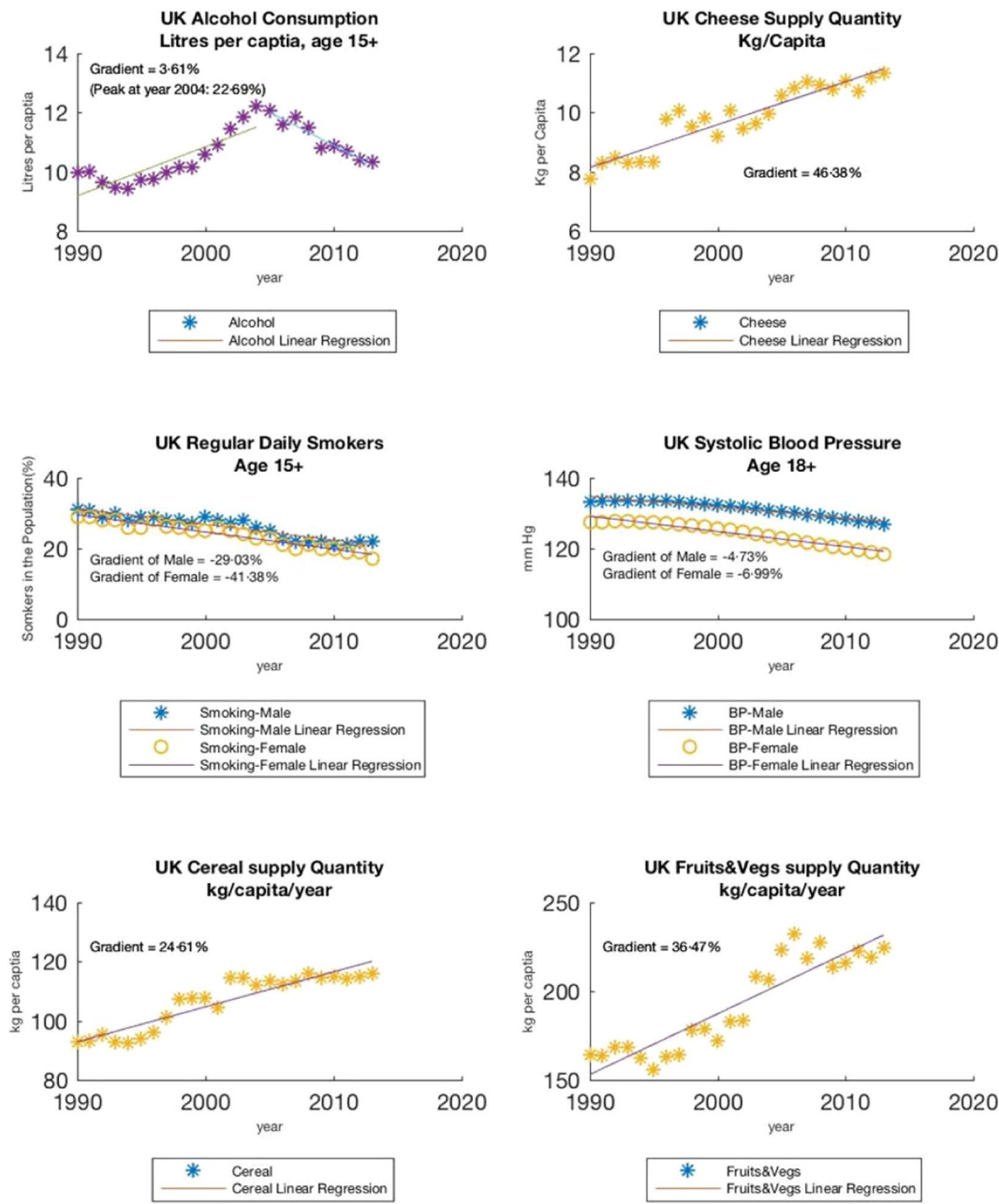

**Figure 2.** Linear plot and linear regression of 6 variables.

*Interdependence between negative and positive factors.* Table 6 shows the relative effects of changes in positive factors over the negative factors. Keeping alcohol, cheese and smoking rates unchanged, a 20% increase in intake of fruit-vegetable was associated with a 3–6% decrease in SBP. A 10% increase in cereal intake was associated with 3% lower impact attributable to SBP. A simultaneous increase of 10% in fruit-vegetable as well as cereal was able to offset the effects of SBP by 6%. These numbers could be evaluated on a patient-by-patient basis using subjective data.

## Discussion

Although CHD deaths have decreased dramatically over the last few decades, it remains a major cause of death worldwide. CHD risk estimation is commonly used in clinical practice to identify patients at risk and targeting known risk factors. In this study, we used a combination of advanced machine-learning techniques to study the mitigating effects of positive lifestyle factors on negative risk factors in a population, using data spanning 1990–2013. Our results show that data based statistical descriptors can be used to estimate the impact of dynamic changes in risk factors and also to quantify the extent to which these risk factors can be ameliorated by positive changes in lifestyle.

|  | T[3] | C[4] | MRREd | MRREl |
|---|---|---|---|---|
| **PCA** | | | | |
| **Male** | 0.9974 | 0.9969 | 0.0547 | 0.0536 |
| **Female** | 0.9911 | 0.9953 | 0.0629 | 0.0540 |
|  | T | C | MRREd | MRREl |
| **GTM** | | | | |
| **Male** | 0.9344 | 0.8833 | 0.2064 | 0.1888 |
| **Female** | 0.9083 | 0.8823 | 0.1999 | 0.2028 |
|  | T | C | MRREd | MRREl |
| **NSC** | | | | |
| **Male** | 0.99969 | 0.9964 | 0.0436 | 0.0433 |
| **Female** | 0.9922 | 0.9958 | 0.0410 | 0.0390 |
|  | T | C | MRREd | MRREl |
| **GPLVM** | | | | |
| **Male** | 0.9052 | 0.9411 | 0.2147 | 0.2130 |
| **Female** | 0.9094 | 0.9401 | 0.2028 | 0.1777 |

**Table 3.** Multivariate Quality Matrix comparing 3 measures.

|  | Male | | Female | |
|---|---|---|---|---|
|  | PC1 | PC2 | PC1 | PC2 |
| **Component Matrix** | | | | |
| Alcohol | 0.49408 | 0.86914 | 0.49412 | 0.86923 |
| Cheese | 0.50206 | 0.26489 | 0.50200 | 0.26940 |
| Smoking | 0.50194 | 0.29280 | 0.50190 | 0.29569 |
| SBP | 0.50188 | 0.29781 | 0.50194 | 0.29058 |

**Table 4.** Component matrix of 4 variables.

|  | Male | | Female | |
|---|---|---|---|---|
|  | PC1 | PC2 | PC1 | PC2 |
| **Component Matrix** | | | | |
| Alcohol | 0.40166 | 0.90956 | 0.40169 | 0.91019 |
| Cheese | 0.40947 | 0.19956 | 0.40944 | 0.20217 |
| Smoking | 0.40941 | 0.22680 | 0.40940 | 0.22812 |
| SBP | 0.40939 | 0.23251 | 0.40943 | 0.22344 |
| Cereals | 0.40974 | 0.10761 | 0.40974 | 0.11118 |
| Fruit-Veg | 0.40975 | 0.12567 | 0.40972 | 0.12790 |

**Table 5.** Component matrix of 6 variables.

| C, F$^\Psi$ | Alcohol | Smoking | Cheese | SBP |
|---|---|---|---|---|
| 1·1, 1·0 | 1 | 1 | 1 | −3·01% |
| 1·1, 1·2 | 1 | 1 | 1 | −9·01% |
| 1·2, 1·3 | 1 | 1 | 1 | −15·02% |
| 1·1, 1·0 | 1 | 1 | −2·04% | 1 |
| 1·1, 1·2 | 1 | 1 | −6·11% | 1 |
| 1·2, 1·3 | 1 | 1 | −10·18% | 1 |

**Table 6.** Numbers calculated from formulae Eq. (1). C, F stands for Cereals and Fruit-Veg.

Over the 23-year period for which data was collected, we observed a steady decline in the rates of CHD across the population. These trends are consistent with those observed in most of the developed world and reflects the effects of a number of health initiatives that have targeted cardio vascular disease that include public health campaign against smoking, blood pressure control and the introduction of statins[25]. The INTERHEART study which looked at potentially modifiable risk factors associated with myocardial infarction identified 9 major risk factors which included amongst other traditional risk factors, cereal and fruit intake, psychosocial health and physical activity highlighting some of these positive indicators of lifestyle have a significant impact on overall risk of CHD[7].

| UK-Male | df | SS | MS | F | Significant F-ρ |
|---|---|---|---|---|---|
| Regression | 1 | 0.049582 | 0.049582 | 7269.007 | 3.18E-29 |
| Residual | 22 | 0.000150 | 6.82E-06 | | |
| Total | 23 | 0.049732 | | | |
| **UK-Female** | **df** | **SS** | **MS** | **F** | **Significant F-ρ** |
| Regression | 1 | 0.041968 | 0.041968 | 5625.284 | 5.28E-28 |
| Residual | 22 | 0.000164 | 7.46E-06 | | |
| Total | 23 | 0.042132 | | | |

**Table 7.** ANOVA test statistic of UK CHD death rate.

In our study, there are notable differences in the trends of different risk factors. While alcohol and cheese consumption increased over time, smoking rates and SBP declined. At the same time, cereal and fruit-vegetable intake markedly increased. These changes reflect the lifestyle alterations in a population over time and while the reduction in smoking and increase intake of cereals and vegetable are encouraging, increases in alcohol and cheese consumption are a concern. Although the changes in CHD rates were comparable between men and women, we observed minor differences in the relative contribution of the different risk factors towards CHD between the genders. While SBP had the biggest impact in men, amongst women it was smoking that had the biggest influence followed by SBP, cheese and alcohol. The differential impact of smoking on CHD risk in women is well known although it is unclear if this is due to the biological differences between genders or a reflection of smoking behaviour[26]. Visualization details are provided as an online appendix.

There are a number of risk factors that contribute to the overall risk of CHD. While all these risk factors have an independent effect on CHD occurrence, the relationship between the risk factors is dynamic and subject to changes over time[27,28]. In the commonly used Framingham risk score a number of risk factors such as age, smoking status, blood pressure and cholesterol levels are used to calculate the risk of CHD[16]. More recently, number of other risk engines have become available. Some of these are disease specific such as the UKPDS risk engine while others[25,29] are more generic and have incorporated new risk factors to enhance the predictability of the overall risk. A common limitation of these risk engines, however, is that they have not explored how CHD risk is altered when life style factors that have beneficial effects on CHD change simultaneously, as is often the case.

The results have all been ANOVA tested, on a factor-by-factor basis, details in the Supplementary. Overall, this leads to a CHD death rate index as in Table 7 above.

Medical opinion presently varies widely between 3–7 fruit-vegetable (FV) portions a day. This study shows that an increase by 20% in FV intake can reduce CHD risk by 3–6%, other risk factors remaining unchanged. If there is a further reduction in smoking by 10%, the same 10% increment in FV consumption will lead to 6% CHD-risk reduction.

Our study has two novel aspects. First, the results presented here are by far the most extensive data modeling (23-year span) done on life-style implications of CHD, where the possibility of simultaneous changes in variables, the so-called "multivariate modeling", has been considered. Second, the conclusions presented have been risk managed over 6 independent machine learning techniques for multivariate analysis, thereby ensuring that the results are probabilistically accurate to the highest possible levels. As is well known, CHD rates are influenced by a number of factors including changes in health care that occur over time. Given that our data was from a defined population who receive health care from the UK-NHS, any changes that may have occurred would be consistent with the trends observed in the whole population.

The key limitation to our modeling is the lack of detailed and reliable data for certain risk factors like age, diabetes status, physical activity or ethnicity. Also, we have only used UK population data, the robustness of which needs to be validated over population surveys involving other countries, a work presently underway.

From a public health perspective, it is now possible to study the long-termed effects of any interventions on known risk factors and measure the effect of those changes over time. More importantly, we present a realistic estimate of how positive lifestyle effects can be encouraged to offset some of the adverse effects of traditional risk factors.

Impact of lifestyle factors on CHD have been previously discussed in health profiling and subsequent medical prognosis. While experience based diagnoses have traditionally differentiated "positive" from "negative" indicators at a qualitative level, our results make quantitative estimates as to how positive factors could "control" detrimental ones, which could lead reduce the prevalence of CHD or delay its onset. Such a population biology-based rescoring of the Framingham mechanism is new and should prove a major diagnostic tool in identifying lifestyle related risk to CHD. A key benefit of this measure will be the CHD risk forecasting at a subjective level using population biology statistics as the median outlier. This study is expected to serve as a key medical and economic incentive of CHD risk assessment[30–35].

### Acknowledgements
AKC and XH are grateful to Darren Flower for careful perusal and comments on the manuscript.

### Author contributions
X.H. and A.K.C. contributed to the study design, data collection, data analysis, data interpretation and writing of the manuscript. X.H. and B.M. contributed to machine learning algorithm. S.B. compared the data modeling results with medical prognosis. G.G. is the forward strategist in connecting the machine learning based research output with tangible patient specific devices/algorithms.

### Competing interests
The authors declare no competing interests.

### Additional information
**Supplementary information** is available for this paper at https://doi.org/10.1038/s41598-020-60786-w.

**Correspondence** and requests for materials should be addressed to A.K.C.

**Reprints and permissions information** is available at www.nature.com/reprints.

**Publisher's note** Springer Nature remains neutral with regard to jurisdictional claims in published maps and institutional affiliations.

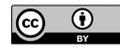 **Open Access** This article is licensed under a Creative Commons Attribution 4.0 International License, which permits use, sharing, adaptation, distribution and reproduction in any medium or format, as long as you give appropriate credit to the original author(s) and the source, provide a link to the Creative Commons license, and indicate if changes were made. The images or other third party material in this article are included in the article's Creative Commons license, unless indicated otherwise in a credit line to the material. If material is not included in the article's Creative Commons license and your intended use is not permitted by statutory regulation or exceeds the permitted use, you will need to obtain permission directly from the copyright holder. To view a copy of this license, visit http://creativecommons.org/licenses/by/4.0/.

© The Author(s) 2020